\documentclass[twocolumn,pra,showpacs,preprintnumbers,amsmath,amssymb]{revtex4}

\usepackage{graphicx}
\usepackage{dcolumn}
\usepackage{bm}
\usepackage{float}
\usepackage{color}

\newcommand{\eps}[0]{\varepsilon}
\newcommand{\rr}[0]{\vec r}
\newcommand{\up}[0]{\uparrow}
\newcommand{\dw}[0]{\downarrow}
\newcommand{\lp}[0]{\left}
\newcommand{\rp}[0]{\right}
\newcommand{\Schr}[0]{Schr\"{o}dinger }
\newcommand{\s}[0]{\sigma}
\newcommand{\sbar}[0]{\bar{\sigma}}

\newcommand{\Hartree}{\textrm{hartree}}
\newcommand{\leqs}[0]{\leqslant}
\newcommand{\ff}[0]{\mbox{\footnotesize{\emph{ff}}}}
\newcommand{\KS}[0]{{K\!S}}

\begin{document}

\title{Higher Ionization Energies of Atoms in Density Functional Theory}

\author{Uri Argaman}
\affiliation{Materials Engineering Department, Ben-Gurion University, Beer Sheva 8410501, Israel}

\author{Eli Kraisler}
\affiliation{Department of Materials and Interfaces, Weizmann Institute of Science, Rehovoth 76100, Israel}

\author{Guy Makov}
\affiliation{Materials Engineering Department, Ben-Gurion University, Beer Sheva 8410501, Israel}

\date{\today}

\begin{abstract}

Density functional theory (DFT) is an exact alternative formulation of quantum mechanics, in which it is possible to calculate the total energy, the spin and the charge density of many-electron systems in the ground state. In practice, it is necessary to use uncontrolled approximations that can mainly be verified against experimental data.  Atoms and ions are simple systems, where the approximations of DFT can be easily tested. We have calculated within DFT the total energies, spin and higher ionization energies of all the ions of elements with $1 \leqs Z \leqs 29$. We find the calculations in close agreement with experiment, with an error of typically less than ca. 1\% for  $1 \leqs Z \leqs 29$. Surprisingly, the error depends on the electronic configuration of the ion in both local spin density approximation (LSDA) and Perdew-Burke-Ernzerhof general gradient approximation (PBE-GGA) and independent of both self-interaction correction (SIC) and relativistic corrections. Larger errors are found for systems in which the spin-spin correlation is significant, which indicates the possible benefit from an orbital-dependent formulation of the correlation energy functional.
\end{abstract}

\maketitle

\section{Introduction.}\label{sec.intro}

Density-functional theory (DFT)~\cite{HK'64,KS'65,PY,DG,Primer,EngelDreizler2011,AM1,Burke12} is the leading theoretical framework for studying the electronic properties of matter. Within the DFT formulation the electron density plays the central role, instead of the many-electron wave function. The approach in DFT is \emph{ab initio}, which means that in principle no experimental data is required to model the system. DFT is currently applied to a variety of many-electron systems~\cite{Martin,Hafner}: along with traditional applications to atoms, molecules and solids, it is used for studying nano-objects, impurities, surfaces, etc. The applications in all these fields are, obviously, interrelated.

The many-body physics of Coulomb-interacting electrons is represented in DFT by the exchange-correlation (xc) energy functional $E_{xc}[n]$. It is non-local, spin-dependent, has non-analytical properties, and its exact form is not known~\cite{DG,Primer,RvL_adv}. There exist, however, many approximations to this functional (e.g.~\cite{VWN'80,PZ'81,PW'92,PW91,PBE'96}), based on numerical results for the homogeneous electron gas~\cite{CepAlder} and asymptotic analytical derivations (see e.g.~\cite{PBE'96} and references therein). An alternative approach to approximating $E_{xc}[n]$ employs Kohn-Sham orbitals in addition to the density, which allows an exact treatment of the exchange (Fock) energy, opening the challenge of formulating a compatible correlation energy, and introducing additional complications~\cite{Primer,Kronik}. The quality of the various approximations can only be determined by testing them on a wide range of systems, in comparison to the experimental data.

Atomic systems are a good class of systems to examine the validity of the approximations introduced into the xc-functional of DFT. The combination of extensive and high-quality experimental data on atoms and ions, which exhibit a rich phenomenology as a function of atomic number and ionization level, together with the relative computational simplicity, which reduces the scope of numerical error, are what makes atomic systems an attractive choice.

DFT allows us to calculate the total energy of atoms or ions. The ionization energy is obtained as the difference between the energy of an ion and the energy of the same ion with one additional electron, which is the neutral atom in the particular case of the first ionization energy. By calculating the first ionization energies and comparing the results to experimental data, the quality of selected xc-approximations (see~\cite{Kraisler} and references therein), the self-interaction correction (SIC)~\cite{Klupfel}, hybrid functionals, e.g.~\cite{HybridChong} and the GW methodology, e.g.~\cite{GW1,GW2,GW3} have been tested in the past. Extending this analysis to higher ionization energies, where the interaction with the nucleus prevails as the atomic number increases and the number of electrons decreases, allows testing the xc-approximations over a much wider range of systems and interaction strengths.

In previous work~\cite{Kraisler}, the first ionization energy of all the atoms with atomic number $Z=1-86$ in both the local-spin density approximation (LSDA)~\cite{PZ'81,VWN'80,PW'92} and Perdew-Burke-Ernzerhof general gradient approximation (PBE-GGA)~\cite{PBE'96} was calculated from the ground state energies of the neutral atoms and first ions. Good overall agreement with experimental data was found, and compatibility with a previous study~\cite{Kot} was maintained in the relevant cases. In addition, it was found that when the electronic configuration was determined \emph{ab initio} rather than empirically, minimizing the energy functional required introducing fractional occupations of the Kohn-Sham orbitals for some atoms and ions. The total spin of the neutral atoms and first ions was also calculated and found to agree with experimental data, except in a few cases. PBE-GGA calculations were found to modestly improve the LSDA calculations.

For the higher ionization energies, there exists an abundance of high accuracy experimental data. Indeed, for atomic numbers $Z=1-29$, all ionization energies have been measured experimentally to good accuracy~\cite{HandChemPhys}. In particular, this allows us to obtain from experiment the total energy by summing the ionization energies.

Several calculations of selected higher ionization energies by both \emph{ab initio} and empirical methods have been reported in the literature. Within the \emph{ab initio} approach are included very accurate quantum chemical methods, which can be applied only to systems with a small number of electrons e.g.~\cite{Chung3,Chung4}, as well as density-functional methods that can be applied to a wide range of systems, in particular to systems with many electrons. In addition, extensive Hartree-Fock (HF) and relativistic Dirac-Fock calculations of ground-state total energies of atoms and ions have been performed~\cite{Clementi,Dirac-Fock}. The higher ionization energies can be obtained from these calculations.

Some simple empirical methods for calculation of the ionization energies have also been reported. An example is Ref.~\cite{Elo}, where the author reports on a simple formula for the ionization energies that depends only on the atomic number and number of electrons. Although this formula is valid only for atoms and ions with 2 and 3 electrons, it predicts ionization energies, including higher ionization energies, to very good accuracy, in these cases.

The total energies of higher ions, in the isoelectronic series $2-18$, were calculated within DFT and compared with non-relativistic estimates of the total energy~\cite{Davidson}. However, as we shall show below, this total energy is a relatively insensitive measure of accuracy compared to the ionization energy, which is a differential quantity and therefore a more sensitive measure of accuracy. To the best of our knowledge, no extensive calculations of the higher ionization energies within density functional theory, and in particular with self-interaction correction (SIC), have been reported.

In the current contribution we present results of self-consistent \emph{ab initio} non-relativistic DFT calculations for atoms and all their ions with $1 \leqs Z \leqs 29$, within the spherical approximation for the density, and the LSDA and PBE-GGA of the xc-energy with and without self-interaction corrections. The objectives of the work are:
(1) to systematically calculate \emph{ab initio} the total energies, ionization energies and spin of atoms and ions within density functional theory;
(2) to explore the effect of the choice of xc-functional between LSDA and PBE-GGA on these calculations;
(3) to explore the effect of the SIC on the error in the calculated ionization energies.
Together these objectives may provide some physical insight into the missing physics in the LSDA and PBE-GGA functionals

The rest of the paper is organized as follows. In Sec.~\ref{sec.theory} we present briefly the theoretical background, in Sec.~\ref{sec.num} the numerical details are given, in Sec.~\ref{sec.results} the ionization energies and total energies for all atoms and ions with $1 \leqs Z \leqs 29$ are reported and compared to experimental data~\cite{HandChemPhys}, Sec.~\ref{sec.discussion} contains a discussion of the errors of the total energies and ionization energies relative to experiment and their dependence on the transition between the Kohn-Sham electronic configurations. Consequently, an orbital-dependent contribution to the correlation functional aimed to better address the spin-spin interactions within the KS system is suggested.

\section{Theory.}\label{sec.theory}

Within the Kohn-Sham (KS) scheme in density-functional theory (DFT)~\cite{KS'65,PY,DG}, it is required to solve a set of one-particle \Schr equations
\begin{equation}
    \lp( - \frac{\hbar^2}{2m_e} \nabla^2 + v_{e\ff,\s}(\rr) \rp) \psi_{i\s} = \eps _{i\s} \psi_{i\s},
\end{equation}
to obtain the eigenvalues $\{\eps _{i\s}\}$ and the orbitals $\{\psi_{i\s}\}$ of the KS system. The KS equations introduce an auxiliary system of non-interacting electrons subject to an effective potential $v_{e\ff}$, chosen so that the density of the KS system equals the density of the interacting system.

The total energy of the interacting system can be expressed in terms of the KS system as
\begin{equation}
    E = T_\KS[n_\up] + T_\KS[n_\dw] + \int v \, n \, d^3r + E_H[n] + E_{xc}[n_\up,n_\dw],
\end{equation}
where $T_\KS$ is the Kohn-Sham kinetic energy functional, $v$ is the external potential, $E_H$ and $E_{xc}$ are the Hartree and the exchange-correlation energies, respectively.

The partial spin-densities $n_\sigma$ have the form:
\begin{equation}\label{ns_EVR}
    n_\s = \sum_{i} g_{i\s} |\psi_{i\s}|^2,
\end{equation}
where $g_{i\s}$ are the occupation numbers, which obey
\begin{equation}\label{gi_EVR}
    g_{i\s} = \left\{
                  \begin{array}{ccc}
                D_{i\s} & : & \eps _{i\s} < \eps _{F\s} \\
                x_{i\s} & : & \eps _{i\s} = \eps _{F\s} \\
                0   & : & \eps _{i\s} > \eps _{F\s}
              \end{array}
           \right. .
\end{equation}
Here $\eps _{i\s}$ is the energy of the $i$th KS level of the $\s$-system, $D_{i\s}$ is the maximal number of electrons that can occupy the $i$-th level, $\eps _{F\s}$ and $x_{i\s} \in [0,D_{i\s}]$ are the energy and the occupation of the highest occupied level(s), which can, in principle, be integral or fractional.

$S$, the $z$-projection of the total spin (referred to in this article as \emph{spin}) is a functional of the partial densities $n_\up,n_\dw$:
\begin{equation}\label{S.def}
    S = \frac{1}{2} \int (n_\up - n_\dw) \, d^3r = \frac{1}{2} \sum_i (g_{i\up} - g_{i\dw}) = \frac{1}{2}(N_\up-N_\dw),
\end{equation}
where $N_\up$ and $N_\dw$ is the number of electrons with spin $\up$ and $\dw$, respectively. Therefore, the spin of the KS system is identical to the spin of the interacting system. However, the occupation numbers $\{g_{i\s} \}$ are internal quantities of the KS systems determined by Eq.~(\ref{gi_EVR}). They are not necessarily equal to the occupation numbers that are available for atoms and ions in the experimental literature~\cite{HandChemPhys}.

In the current work the physical quantities obtained in DFT calculations are the partial densities $n_\up$ and $n_\dw$ and $E(Z,N)$ -- the total energy of a system with atomic number $Z$ and $N$ electrons. The ionization energy
\begin{equation}\label{I.def}
I(Z,N) = E(Z,N-1)-E(Z,N),
\end{equation}
and the spin $S$ are derived from these quantities (see Eq.~(\ref{S.def})).

The problem of self-interaction in DFT is well-known~\cite{PZ'81}, and it becomes obvious when considering one-electron systems. In this case, the standard energy functional includes the following contributions from electron-electron interactions: electrostatic energy, exchange energy and correlation energy, all three of which, of course, do not exist in one-electron systems and should cancel out for the exact xc-functional. However, in any approximate functional these three terms do not cancel out and a residual self-interaction error remains. We note that this problem does not exist in the Hartree-Fock method because the self-interaction electrostatic energy exactly cancels the self-interaction exchange energy and no correlation term exists.

In the one-electron case, a self-interaction correction is easily applied by removing all electron-electron interactions. In a more general case, where the systems considered include more than one electron, implementation of a SIC becomes more complex. A first version of SIC was proposed by Fermi and Amaldi in 1934~\cite{Fermi}, within the Thomas-Fermi theory. Later, Perdew and Zunger~\cite{PZ'81} formulated a SIC by requiring that:
\begin{equation}\label{SIC1}
E_H[n_{i\sigma}]+E_{xc}[n_{i\sigma},0]=0,
\end{equation}
where $n_{i\sigma}=|\psi_{i\s}|^2$ is the density of the $i\sigma$-th orbital. In a more detailed form:
\begin{equation}\label{SIC2}
E_H[n_{i\sigma}]+E_{x}[n_{i\sigma},0]=0
\end{equation}
and
\begin{equation}\label{SIC3}
E_{c}[n_{i\sigma},0]=0
\end{equation}
So, in a given approximation of $E_{xc}$, the excess self-interaction can be corrected by replacing the xc-functional $E_{xc}$ with
\begin{equation}\label{SIC4}
E_{xc}^{SIC}[n_\up,n_\dw]=E_{xc}[n_\up,n_\dw]-\sum_{i\sigma}(E_H[n_{i\sigma}]+E_{xc}[n_{i\sigma},0])
\end{equation}

In principle, in the Perdew-Zunger SIC approach, one must correct the potential and the density, as well as the energy functional. However, due to the variational nature of the problem, the total energy is independent, to the first order, of changes in the density and the self-interaction correction may be approximated by correcting just the energy. Therefore, in the current work we apply the SIC corrections to the energy values only, after performing the standard self-consistent DFT calculation.

\section{Numerical Methods.} \label{sec.num}

As in earlier studies of atomic systems~\cite{MJW,Kot,EngelOathDrei,Kraisler}, in the current work the density was approximated by its spherical average $n_\s (r) = (4\pi)^{-1} \int_0^{2\pi} n_\s(\rr) d\Omega$. Since $v$ is spherical, the effective potentials are spherical, too. This approximation reduces a three-dimensional problem to a one-dimensional problem. With a spherical potential, the wave functions take the form $\psi_{nlm\s}(\rr) = R_{nl\s}(r) Y_{lm}(\theta,\phi)$, where $Y_{lm}(\theta,\phi)$ are the spherical harmonics and $R_{nl\s}(r)$ are radial wave functions that obey the following differential equation:
\begin{equation} \label{Schr.radial.eq}
R_{nl\s}'' + \frac{2}{r} R_{nl\s}' + \lp( 2 \frac{m_e}{\hbar^2}(\eps _{nl\s} - v_{e\ff,\s}(r))- \frac{l(l+1)}{r^2} \rp) R_{nl\s} = 0.
\end{equation}
Since the energy levels do not depend on $m$, $D_{nl\s}=2l+1$ where $D_{nl\s}=2l+1$ is the degeneracy of the $nl\s$-th level.

The electronic configuration in the KS system is determined \emph{ab initio}, by requiring the total energy to be a minimum for any fixed number of electrons. Relying on a previous study~\cite{Kraisler}, we expect systems from the $s$- and $p$-blocks of the Periodic Table to follow Hund's rule~\cite{LL3}. In the $d$-block, however, i.e.\ for $21 \leqs Z \leqs 29$, where the $3d$- and the $4s$-levels are rather close, we explicitly checked for a possible degeneracy leading to fractional occupation.

In all calculations, a high numerical convergence of $10 \:\:\mu\Hartree$ for the total energy was obtained. To assure the desired accuracy in energy, Eq.~(\ref{Schr.radial.eq}) was solved on a logarithmic grid with 5,000 points, on the interval $(e^{-a}/Z,L)$ in Bohr radii, with $a=13$ and $L=25$ for both LSDA and PBE-GGA.

\section{Results}\label{sec.results}

The ground-state energy, the spin and the KS electronic configuration have been obtained for all the atoms and ions with atomic number $Z=1-29$, within both LSDA and PBE-GGA. Self-interaction corrections to the total energy have been also calculated for all atoms and ions with $Z=1-18$. The numerical uncertainty was estimated to be less than  $10 \:\:\mu\Hartree$ for the total energy (see above). Therefore we assume below that the entire difference between the calculated and measured ionization energies is due to physical approximations in the energy functional.

All the experimental ionization energies with which comparison is made are taken from the Handbook of Chemistry and Physics~\cite{HandChemPhys}, except for the ionization energy of Fe$^{23+}$, which was obtained following the analysis of Ref.~\cite{Reader}.

\subsection{Total energy}
We first focus our interest on the relative error in the total ground-state energy determined for a system with $N$ electrons and atomic number $Z$. The ground-state total energy of such a system can be calculated from experimental data as follows:
\begin{equation}
E_{tot}^{exp} (Z,N) = - \sum_{M=1}^N  I^{exp}(Z,M)
\end{equation}
where $I^{exp}(Z,M)$ is the experimental ionization energy of the atom or ion with atomic number $Z$ and number of electrons $M$. The relative error is then defined as
\begin{equation} \label{Error}
    \Delta E= \frac{E_{tot}^{calc}-E_{tot}^{exp}}{E_{tot}^{exp}},
\end{equation}
where $E_{tot}^{calc}$ is the calculated ground-state energy. The relative error for the ionization energy is defined in a similar manner.

\begin{figure*}
    \centering
    \includegraphics[trim=0mm 0mm 0mm 0mm,scale=0.7]{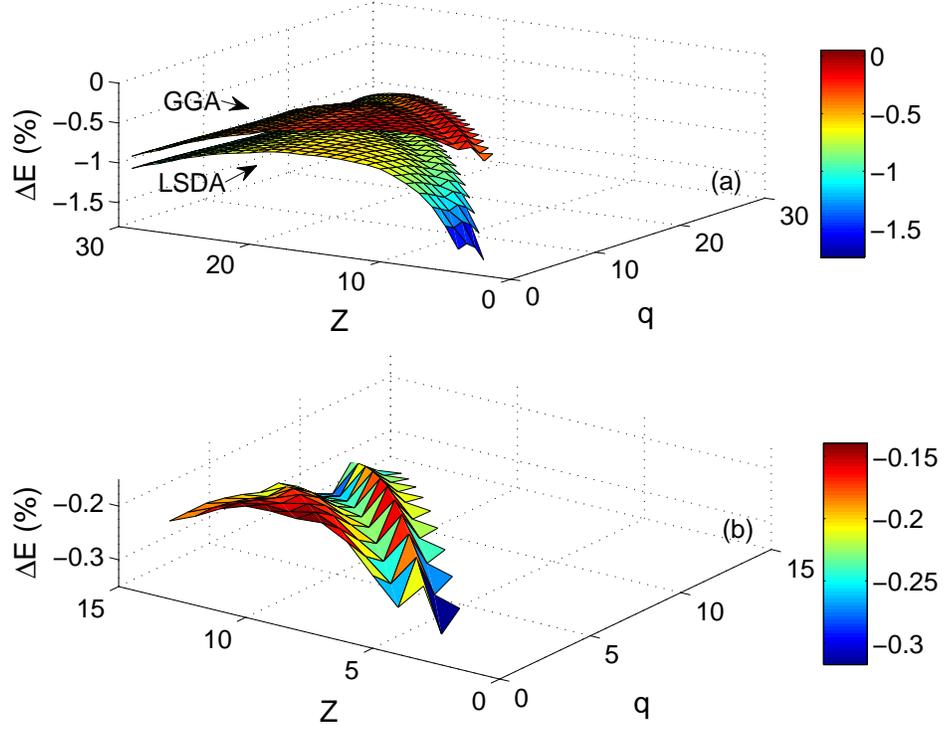}\\
    \caption{(Color online) Relative error in the total energy as a function of the atomic number, $Z$, and the ionic charge, $q$: (a) LSDA and PBE-GGA (b) Enlargement of a section of the PBE-GGA error surface exhibiting undulatory structure.}\label{totalEnergy3d}
\end{figure*}

Fig. ~\ref{totalEnergy3d} presents the relative error in the total energy of atoms and ions as a function of $N$ and $Z$, for $Z=1-29$, obtained within the LSDA and PBE-GGA. It is immediately evident from the figure that the PBE-GGA shows better agreement with the experimental data than the LSDA, as the GGA error surface lies closer to the zero error plane for any value of $(Z,N)$. The error is negative, i.e.\ the total energy estimated by DFT is less than that obtained experimentally, and in general the error is smoothly dependent upon $(Z,N)$. The magnitude of the relative error is seen to increase for small $N$, as would be expected for DFT, and for large values of $Z$ due to relativistic effects (for an overview on relativistic DFT, see e.g.~\cite{DG,Primer,EngelDreizler2011,Autschbach12,EngelOathDrei}). We note that there is a deviation from the smooth dependence of the error for ions with four electrons. Close inspection, performed in panel (b) of Fig.~\ref{totalEnergy3d}, shows that the error surface is in fact not entirely smooth over the $(Z,N)$ plane and exhibits several oscillations.

\begin{figure}
    \centering
    \includegraphics[trim=0mm 0mm 0mm 0mm,scale=0.5]{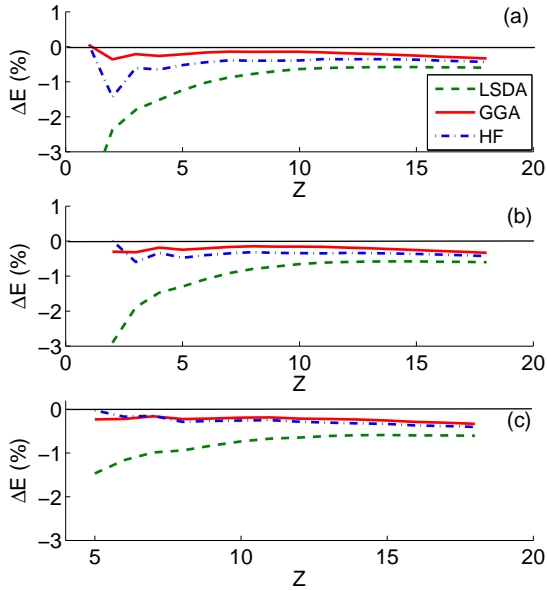}\\
    \caption{(Color online) Relative error $\Delta E$ in the total energy in LSDA, PBE-GGA and HF~\cite{Clementi} approximations: (a) neutral atoms, (b) +1 charged ions and (c) +4 charged ions.}\label{totalEnergy1}
\end{figure}

\begin{figure}
    \centering
    \includegraphics[trim=0mm 0mm 0mm 0mm,scale=0.5]{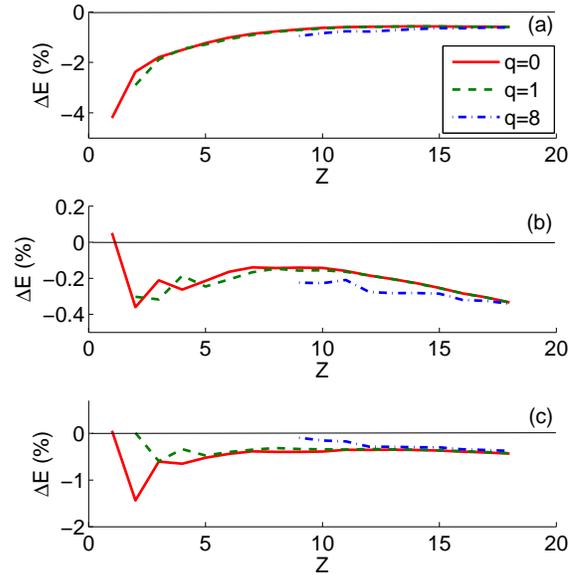}\\
    \caption{(Color online) Relative error $\Delta E$ in the total energy with a different ionic charge, $q=Z-N$. (a) LSDA, (b) PBE-GGA (this work) and (c) HF (Ref.~\cite{Clementi})}\label{totalEnergy2}
\end{figure}

We compare our LSDA and PBE-GGA calculations with HF results obtained by Clementi and Roetti~\cite{Clementi}. It can be seen in Fig.~\ref{totalEnergy1} that the errors in the calculated total energy within all approximations decrease initially with atomic number. The errors in the HF approximation approach the error of the PBE-GGA faster than those of the LSDA. At higher $Z$ the relativistic effects increase and are not included in any of the approximations above, so that the relativistic error is the same in all approximations and is expected to become dominant at higher values of $Z$.

In Fig.~\ref{totalEnergy2} we see that the relative error in the total energy within a given xc-approximation varies slowly with the nucleus charge, $Z$, and is almost independent of the number of electrons represented by the net ionic charge , $q=Z-N$. From this result, together with the dominant contribution of the core electrons to the total energy, we can conclude that the dominant contribution to $\Delta E$ arises from errors associated with the core electrons. Therefore, it is of interest to analyze the error in the ionization energy, which is obtained as a difference between total energies, and in which therefore the errors originating in the core electrons cancel, to a large extent.

\subsection{Ionization energies}

All the higher ionization energies (IE) for systems with $Z=1-29$ were calculated within both LSDA and PBE-GGA. The typical error is of the order of $1\%$ relative to experiment, except for several low ionization energies. Both LSDA and PBE-GGA appear to be satisfactory choices of the xc-functional for calculating the higher ionization energies as discussed below. In the context of ionization energy, the notation $N$ is denotes the number of electrons of the atom or ion before ionization.

\begin{figure*}
    \centering
    \includegraphics[trim=0mm 0mm 0mm 0mm,scale=0.7]{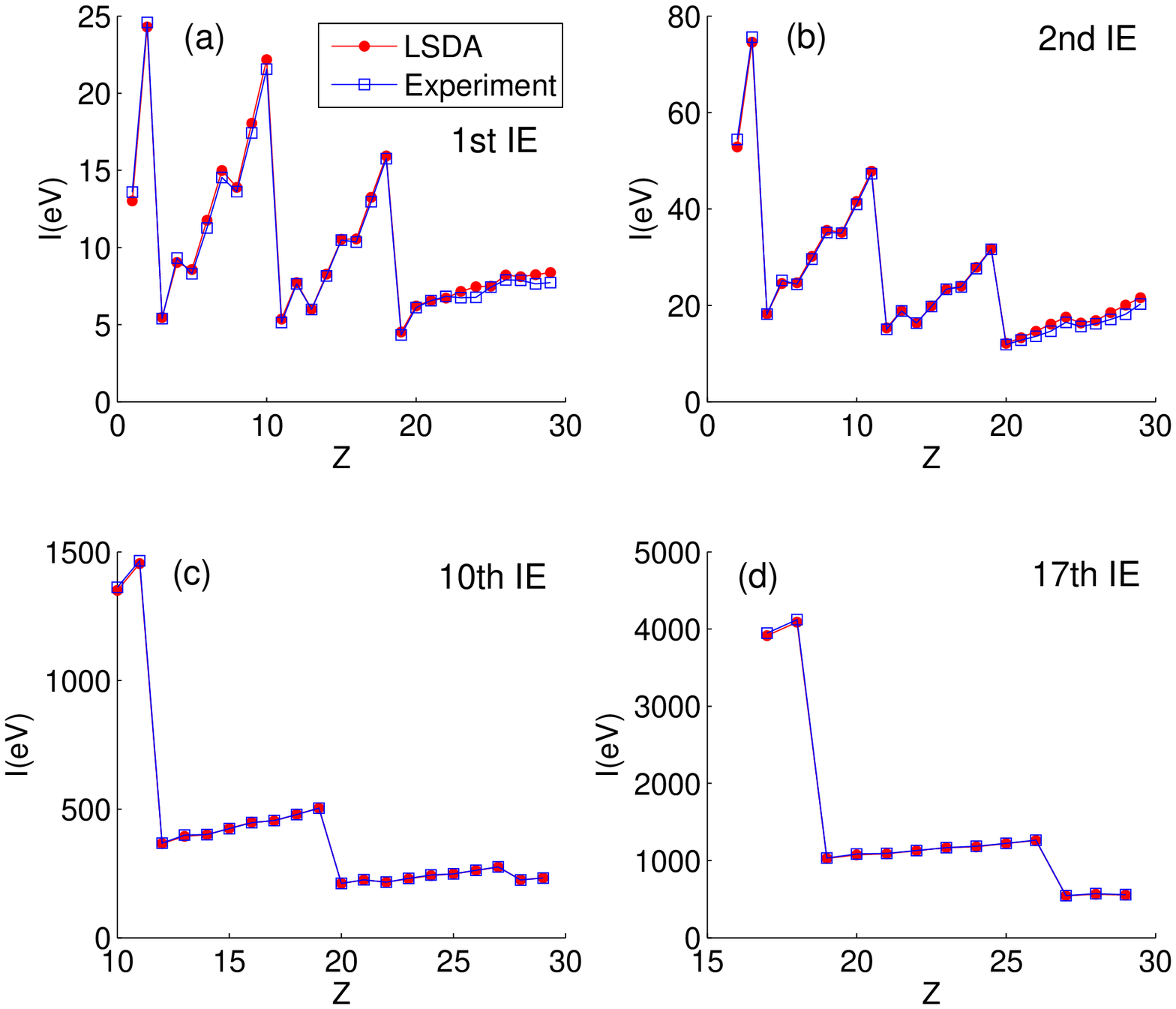}\\
    \caption{(Color online) Higher ionization energies as a function of atomic number, calculated within the LSDA and compared with experimental data. (a) first, (b) second, (c) tenth and (d) 17th ionization energies.}\label{hie}
\end{figure*}

The 1st, 2nd, 10th and 17th ionization energies as a function of atomic number are calculated within the LSDA are compared with experimental ionization energies~\cite{HandChemPhys} in Fig.~\ref{hie}. The results for the first ionization energy are the same as obtained in~\cite{Kraisler}. At higher ionization energies the detailed structure of the atomic sub-shells and orbitals diminishes, and the $Z$-dependence is dominated by the shell structure. Similar results are obtained in the PBE-GGA.

It can also be seen from Fig.~\ref{hie} that when the $p$-shell is half-filled, the first ionization energy decreases in both theory and experiment. For the second ionization energy, this behavior is also observed. At higher ionization energies this effect disappears because of the stronger interaction with the nucleus.

The magnitude of the ionization energy increases rapidly with ionic charge making comparison of errors difficult. Therefore, further analysis should focus on the relative, rather than the absolute errors, as defined in Eq.~(\ref{Error}).

\begin{figure*}
    \centering
    \includegraphics[trim=0mm 0mm 0mm 0mm,scale=0.7]{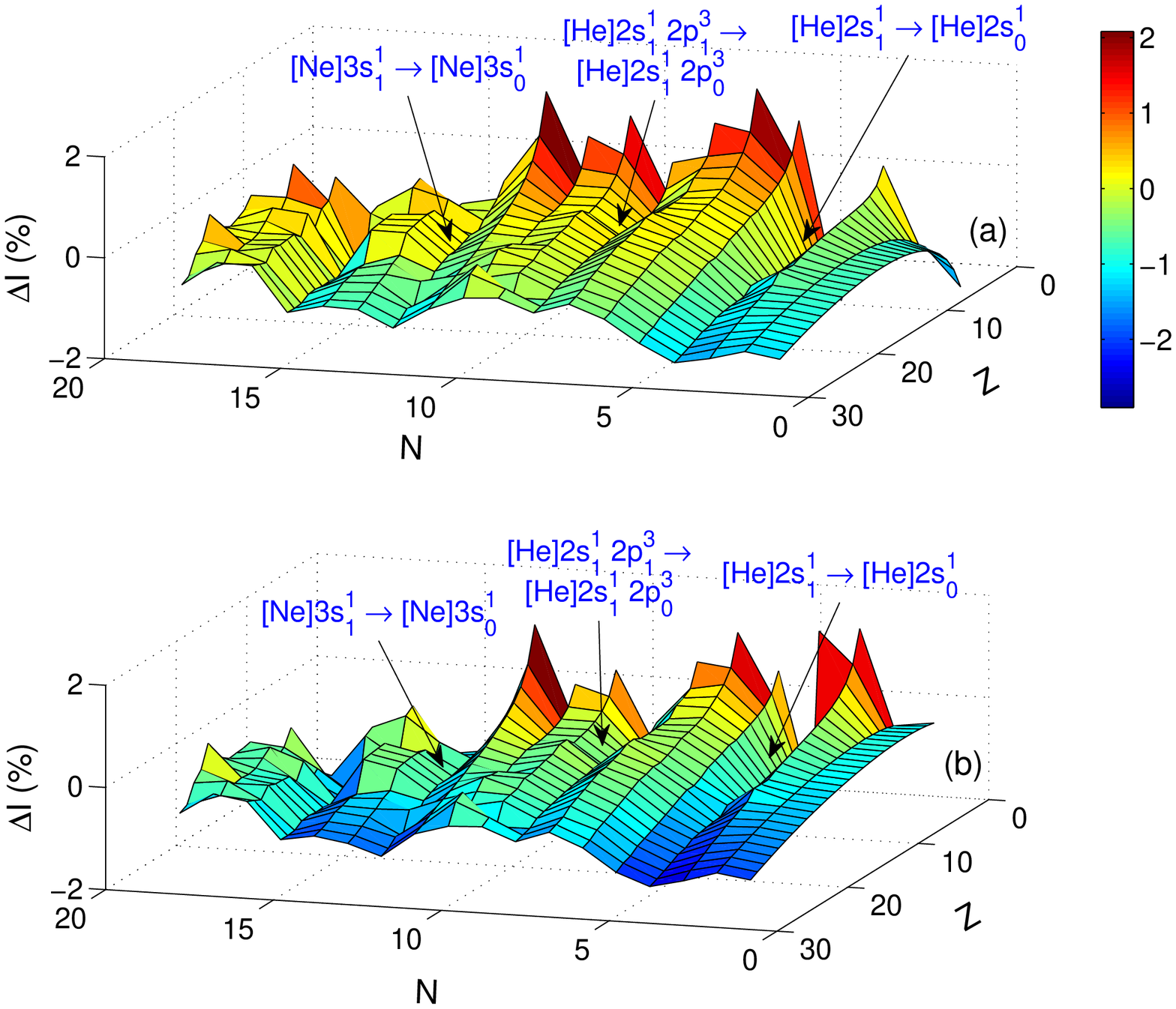}\\
    \caption{(Color online) Relative error surfaces of the ionization energy in (a) LSDA and (b) PBE-GGA. The error surfaces are not smooth and exhibit structural dips at transitions between specific electronic configurations in both PBE-GGA and LSDA approximations.}\label{ldagga}
\end{figure*}

\begin{figure}
    \centering
    \includegraphics[trim=0mm 0mm 0mm 0mm,scale=0.40]{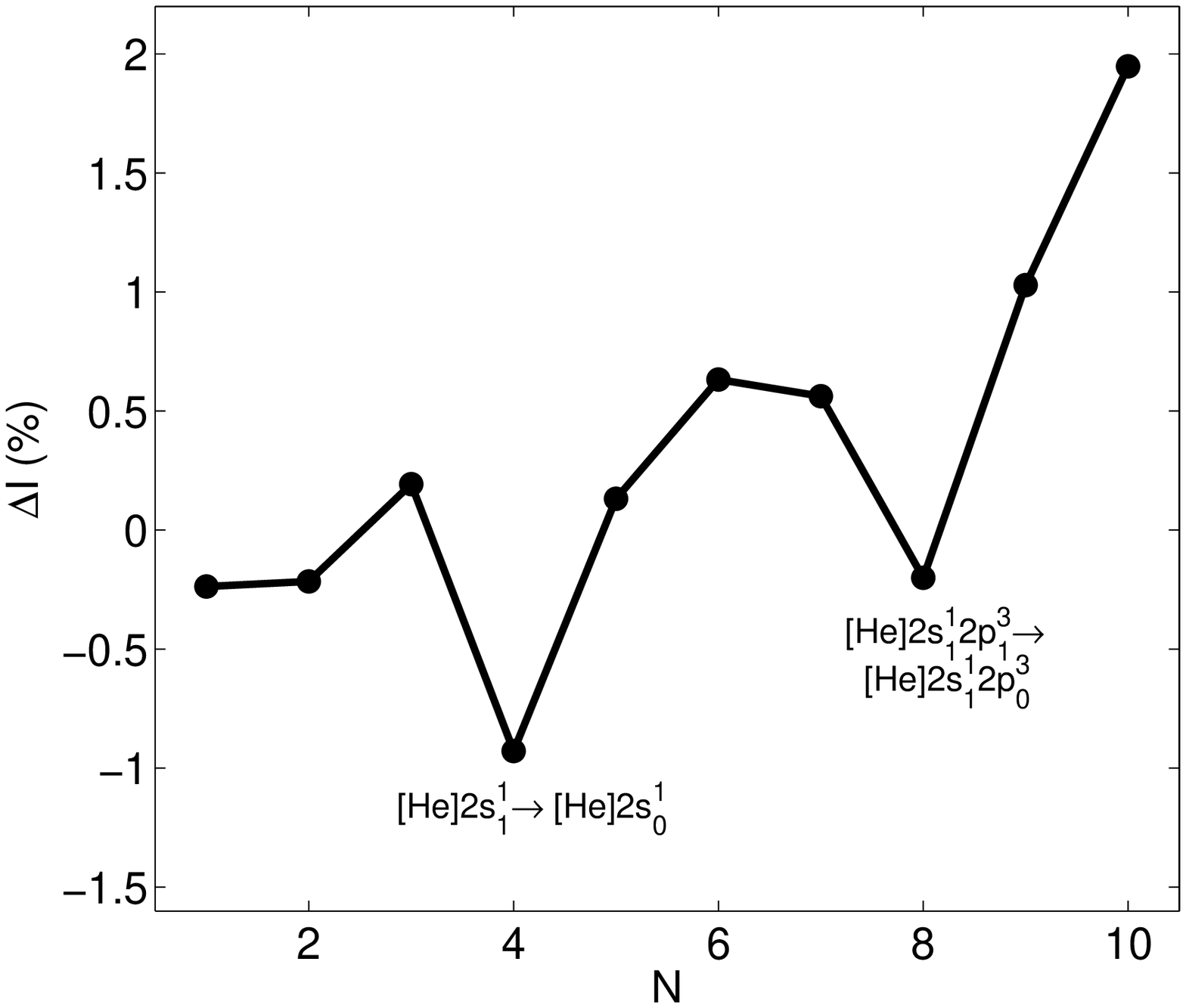}\\
    \caption{Relative error of the ionization energies in the PBE-GGA calculations in constant atomic number $Z=10$.} \label{ggaerror}
\end{figure}

In Fig.~\ref{ldagga} we present the relative error in the ionization energies of atoms and ions with $Z=1-29$, calculated in both the LSDA and PBE-GGA as a function of $Z$ and $N$. It can be seen that the relative error has a certain pattern: In the background of the figure, which corresponds to the first ionization energies, the errors are large, decreasing to the foreground, which corresponds to higher ionization energies. Furthermore, the decrease is not uniform. Instead, we see dips, i.e. large negative errors across the $(Z,N)$ plane for certain values of $N$, which correspond to transitions between specific electronic configurations. For example, a large negative error is found in the ionization energy from the 4-electron configuration $1s^1_1 2s^1_1$ to the 3-electron configuration $1s^1_1 2s^1_0$, correlated with the structure seen in the error of the total energy for 4-electron ions.

In Fig.~\ref{ggaerror} we can see a cut in the error surface of the ionization energies along the $(Z,N)$ plane at constant atomic number so that every point on the graph represents a different configuration. In this figure it is easy to see that the increased magnitude of error occurs at specific configurations namely $2s^1_1 \rightarrow 2s^1_0$, $2p^3_1 \rightarrow 2p^3_0$. This finding remains unchanged between the LSDA and PBE-GGA.

\begin{figure}
    \centering
    \includegraphics[trim=0mm 0mm 0mm 0mm,scale=0.4]{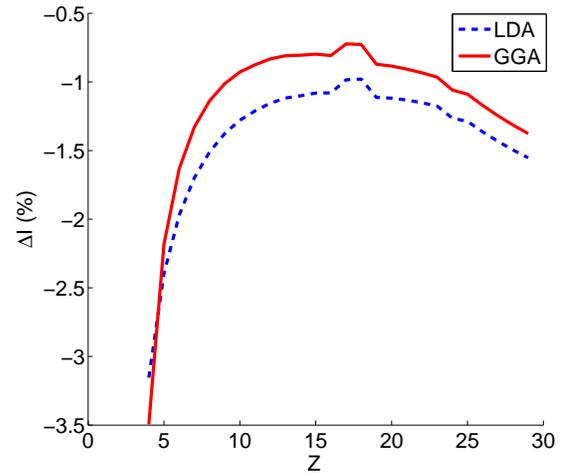}\\
    \caption{(Color online) Relative error in the $1s^1_1 2s^1_1 \rightarrow 1s^1_1 2s^1_0$ ionization energy in LSDA and PBE-GGA.} \label{1s2}
\end{figure}

The effect of the atomic number on the error for a fixed electronic transition from four to three electrons is presented in Fig.~\ref{1s2} for both the LSDA and the PBE-GGA. The dependence of the error on $Z$ is similar in both approximations, so that, also here, the PBE-GGA does not make a qualitatively difference relative to the LSDA. Furthermore, we see a rise in the relative error for high values of $Z$, due to relativistic effects.

In passing, we wish to draw the attention to the deviation of the curves in Fig.~\ref{1s2} from the trend around $Z=17-18$. The same pattern in the error was discovered by Chung~\emph{et al.}~\cite{Chung4} using a completely different method. The magnitude of the deviation from experiment is similar to our PBE-GGA and LSDA calculations, of the order of 0.1\%. This deviation might be explained by assuming that there is a small error in the experimental data, not unlike the case of Fe$^{23+}$~\cite{Reader}.

\subsection{Self-interaction correction}

\begin{figure}
    \centering
    \includegraphics[trim=0mm 0mm 0mm 0mm,scale=0.4]{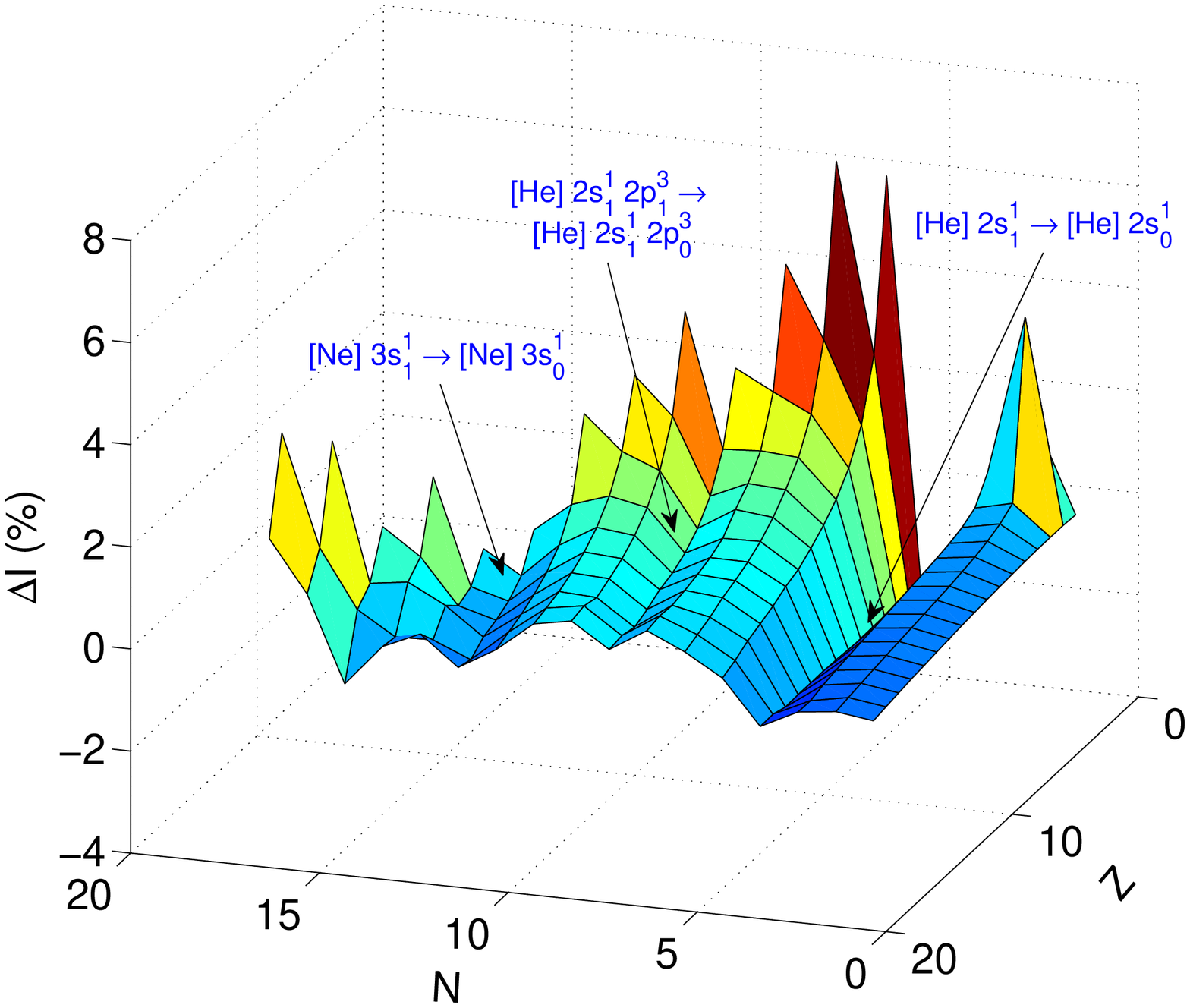}\\
    \caption{(Color online) Relative error in the ionization energy, obtained for the LSDA with SIC. The dips in the error surface retain after SIC is applied.}\label{SIC}
\end{figure}

Further insight into the origin of the pattern in the error surface presented in Fig.~\ref{ldagga} might be obtained by considering the effect of self-interaction corrections. We have calculated the higher ionization energies with a first order SIC to the total energy, as explained in Sec.~\ref{sec.theory}, for $Z=1-18$. The relative error surface calculated with SIC across the $(Z,N)$ plane is shown in Fig.~\ref{SIC} similar to Fig.~\ref{ldagga}. From Fig.~\ref{SIC} it can be seen that the agreement between the calculated ionization energy and experiment is not always improved by SIC (despite an improvement in total energy values as was previously observed in Ref.~\cite{Klupfel}). In particular, the pattern of the relative error in the ionization energy observed in LSDA and PBE-GGA remains essentially unchanged after performing the self-interaction correction.

\subsection{\emph{Ab initio} configuration and spin}

In a previous work~\cite{Kraisler}, it was found that for all atoms and first ions with atomic number $Z=1-29$ the calculated total spins agree with the experiment, except for the Ti and V atoms. In addition, Fe and Co atoms and Sc and Ti first ions were found to have fractional occupation numbers in the KS system. Moreover, the Ni atom was found to have a KS electronic configuration that is different from the reported experimental (empirical) configuration~\cite{HandChemPhys}, although the total spin is predicted correctly. As is discussed in~\cite{Kraisler}, the Kohn-Sham configuration does not have to be the same as the experimental configuration.

For higher ions (ionization level greater than 1) with atomic number $Z=1-29$, we have found that all the KS electronic configurations, obtained both with the LSDA and the PBE-GGA fit the reported experimental data. Additional cases of fractional occupation were not found. This finding is not unexpected, because of the stronger interaction with the nucleus in higher ions which results in a larger separation between the $3d$ and $4s$ KS levels. This reduces the possibility of their fractional occupation and therefore precludes the appearance of ensemble-state solutions.

\section{Discussion.}\label{sec.discussion}
We have seen that the error in the ionization energy relative to experiment is of the order of $1\%$ and that it does not evolve uniformly across the $(Z,N)$ plane. Instead, at certain electronic configuration transitions it experiences dips -- a sharp increase in the absolute value of $\Delta I$, while it becomes negative. This occurs at the transitions $2s^1_1 \rightarrow 2s^1_0$, $2p^3_1 \rightarrow 2p^3_0$ and $3s^1_1 \rightarrow 3s^1_0$, for both the LSDA and the PBE-GGA, as can be seen on Fig.~\ref{ldagga}.

\begin{figure}
    \centering
    \includegraphics[trim=0mm 0mm 0mm 0mm,scale=0.4]{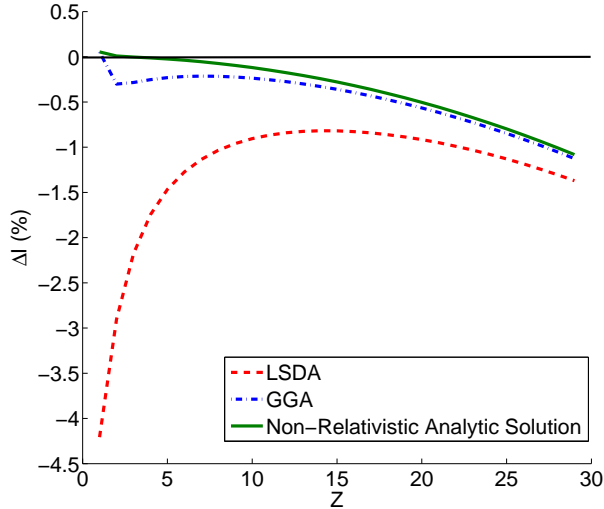}\\
    \caption{(Color online) Relative error in the LSDA and PBE-GGA calculations and relative error in the non-relativistic analytic solution for hydrogen like-ions.}\label{hla}
\end{figure}

The error in the ionization energy of ions can originate from several sources, and we discuss them below. One possible source of error is relativistic effects. In an arbitrary atom or ion, it is difficult to isolate the error introduced by using non-relativistic DFT. However, in the case of one-electron ions, we obtain the relativistic contribution to the error exactly: it equals the difference between the ground-state energy obtained via an analytical solution to the one-electron non-relativistic Schr\"{o}dinger equation and the experimental ionization energy. In Fig.~\ref{hla} one observes the deviation of the analytical solution from the experimental data, which increases with atomic number, as expected. On the same figure we plot the relative errors for one-electron ions obtained with LSDA and PBE-GGA calculations. We recall that in this specific case the ionization energy equals minus the ground state energy in the $1s^1_0$ configuration, and that the error in DFT result comes from two sources: self-interaction and relativistic effects. Comparing the curves in Fig.~\ref{hla} we see that the PBE-GGA solution converges to the analytical solution faster than the LSDA solution. That is, the energy of one-electron ions is represented more accurately in PBE-GGA. Surprisingly, we see that for one-electron systems, the relativistic contribution to the error dominates in the LSDA already for $Z>20$ and in the PBE-GGA even for $Z>10$. This is contrary to the view that relativistic effects are important only for heavy atoms.

However, the non-uniform pattern in $\Delta I$ is not of a relativistic source: from the work of Chung~\emph{et al.}~\cite{Chung3,Chung4} one can deduce that the difference between relativistic and non-relativistic ionization energies for 4-electron systems with a low $Z$, for which we observe the first dip, is much smaller than the magnitude of the dip.

Although the self-interaction correction improves the total energy obtained, albeit overcorrecting somewhat, we see from Fig.~\ref{SIC} that the self-interaction correction does not remove the aforementioned pattern in the ionization energy error surface. Moreover, the influence of the spherical approximation is also ruled out, at least for the transitions $2s^1_1 \rightarrow 2s^1_0$ and $3s^1_1 \rightarrow 3s^1_0$, because both ions are then completely spherical.

Therefore, we reach the conclusion that \emph{neither the locality of the xc-energy, nor the self-interaction, nor relativistic errors, nor the spherical approximation to the density are the reason for the non-uniform pattern in the error $\Delta I (Z,N)$}.

In search of a possible explanation, we note that the dips in the error $\Delta I$ occur when the final configuration has a fully polarized (say, $\up$) subshell at the highest level, while the initial configuration has an additional electron in the other (say, $\dw$) subshell. It is therefore reasonable to assume that in these cases the spin-spin interaction is rather significant, and probably is not described accurately enough by common xc-functionals.

\begin{figure}
    \centering
    \includegraphics[trim=0mm 0mm 0mm 0mm,scale=0.4]{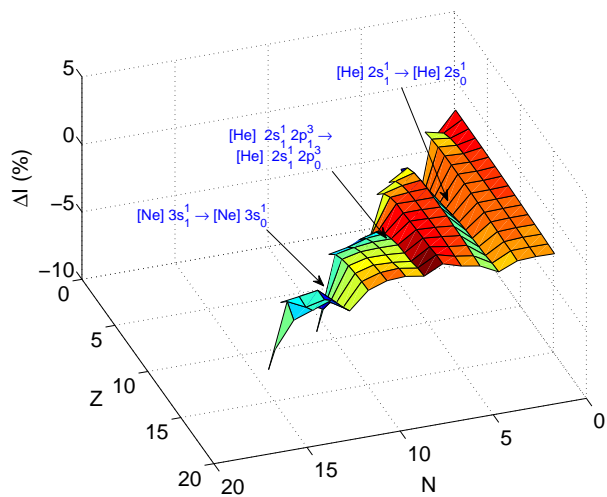}\\
    \caption{(Color online) Relative error in the ionization energy in the HF method.} \label{HF3d}
\end{figure}

\begin{figure}
    \centering
    \includegraphics[trim=0mm 0mm 0mm 0mm,scale=0.45]{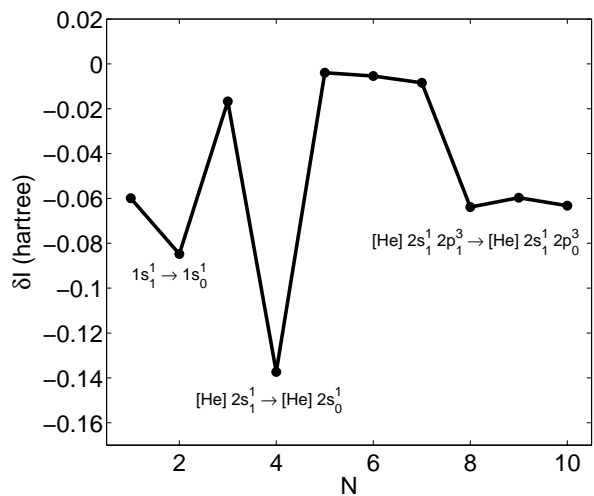}\\
    \caption{Absolute error in the ionization energies, $\delta I$, in the HF method with $Z=10$.} \label{HF2d}
\end{figure}

To further analyze the dependence of the error $\Delta I$ on the electronic configuration transition, we determine the ionization energies in the Hartree-Fock approximation from the total energies obtained by Clementi and Roetti~\cite{Clementi} for atoms and ions. The relative error in the ionization energy as a function of $Z$ and $N$ is presented in Fig.~\ref{HF3d}. From this figure it can be seen that $\Delta I$ possesses the same pattern as in the LSDA and PBE-GGA cases. Furthermore, Fig.~\ref{HF2d} presents the absolute error, $\delta I = I_{HF} - I^{exp}$ -- the difference between the HF and the experimental ionization energies, for $Z=10$. Here we clearly see the dip that corresponds to the transition $2s^1_1 \rightarrow 2s^1_0$, followed by a plateau of almost zero error for $N=5,6,7$, i.e.\ when filling the $\up$-$p$-subshell of the ion, and another plateau of an error of $\sim 0.06 \:\: \Hartree$ for $N=8,9,10$, i.e. when filling the $\dw$-$p$-subshell of the ion.

Recall that the HF method does not include correlation effects, while treating the exchange exactly. Therefore the absolute error in the ionization energy, $\delta I$, may be considered as a rough approximation to the correlation energy associated with the last electron. From Figs.~\ref{HF3d} and~\ref{HF2d} we learn that the correlation energy of the last electron depends strongly on the orbital and its spin occupancy, and only weakly on the densities $n_\up$ and $n_\dw$. In particular, the correlation energy of the last electron is significantly increased if it is in a doubly occupied orbital.

This analysis leads us to a conclusion that a local (or a semi-local) density-based correlation energy is insufficient to accurately obtain the correct total and ionization energies for ions, due to an insufficiently accurate description of the  spin-spin interaction. Alternative perturbation based orbital approaches to the correlation energy, e.g. Ref.~\cite{mori} are also inadequate as they focus on interaction with unoccupied orbitals. Therefore, we suggest considering a contribution to the correlation energy functional in terms of the Kohn-Sham orbitals~\cite{Kronik} a conclusion similar to that obtained from different considerations in a study of self-interaction corrections in atomic systems by Kl\"upfel ~\emph{et al.}~\cite{Klupfel}. In the limit of a fully delocalized electron gas this formulation should reduce to the LSDA. Let us introduce
\begin{equation}
    Q_{i\s,j\tau} = \int \psi_{i\s}^*(\rr) \hat{A}_{i\s,j\tau} \psi_{j\tau}(\rr) d^3r
\end{equation}
as an interaction between KS orbitals characterized by the quantum numbers $(i,\s)$ and $(j,\tau)$, correspondingly, via an undefined operator $\hat A_{i\s,j\tau}$. Next, inspired by Fig.~\ref{HF2d}, we suggest that every orbital $(i,\s)$ will interact only with the orbital that has the same spatial quantum numbers and a different spin. Mathematically, this requirement can be expressed as $\hat{A}_{i\s,j\tau} = \hat{A}_{i\s} \delta_{ij} (1- \delta_{\s \tau})$. Because of the term $(1- \delta_{\s \tau})$ we do not include any correction to the interaction of the orbital with itself. Then, the proposed contribution to the correlation energy equals
\begin{eqnarray}
    E_c^{corr} [\{ \psi_{i\s} \}] &=& \frac{1}{2} \sum_{i\s} \sum_{j \tau} g_{i\s} g_{j \tau} Q_{i\s,j\tau}   \\
		&=& \frac{1}{2} \sum_{i\s} g_{i\s} g_{i \sbar} \int \psi_{i\s}^*(\rr) \hat{A}_{i\s} \psi_{i\sbar}(\rr) d^3r, \nonumber
\end{eqnarray}
where $\sbar = \dw$ when $\s = \up$, or vice versa and $g_{i\s}$ are the occupation numbers defined in Eq.~(\ref{gi_EVR}). The exact form of the term $\hat{A}_{i\sigma}$ is currently unknown. Possibly relevant work in this direction may be the application of GW methodology to the calculations of atomic systems, as, for example, in  Refs.~\cite{GW1,GW2}. We expect, though, that such an additional term, which includes interaction between orbitals of opposite spins, will be able to remedy the errors in the ionization energies discussed above.

\section{Summary}\label{sec.summary}
We have calculated the total energies of all ions and atoms with $1 \leqs Z \leqs 29$ in the LSDA and PBE-GGA in density-functional theory. We find that the absolute value of the relative error in the total energy decreases with increasing $Z$ in the non-relativistic regime. The lowest error is obtained in the PBE-GGA, followed by the HF approximation and finally the largest error is found in the LSDA. From the total energies it is possible to obtain the ionization energies as finite differences and compare these directly to experiment. It is found that the ionization energies are typically reproduced to an accuracy of better than 1\% with the error decreasing and becoming (more) negative as the atomic number increases. At low atomic numbers, for a given number of electrons, the error is apparently dominated by electron-electron interactions whereas at high atomic numbers the electron-nucleus interaction dominates and introduces a relativistic error. It was found that the error in the ionization energy strongly depends on the configuration transition in the LSDA, PBE-GGA, LSDA-SIC and HF calculations. As a result, employment of an orbital-dependent correlation energy that includes interaction between opposite spin channels was proposed.

\bibliography{atoms_ions}

\end{document}